\documentclass[12pt]{article}
\usepackage{epsfig,textpos,url}
\usepackage{lineno}
\begin{document}
%\linenumbers
\title{Conceptual design of a mid-energy spin rotator\\
  for continuous-wave operation based on a\\
  compact multi-pi rosetta magnet}
\author{Volker Ziemann, Jefferson Lab}
\date{July 31, 2026}
\maketitle
%%%%%%%%%%%%%%%%%%%%%%%%%%%%%%%%%%%%%%%%%%%%%%%%%%%%%%%%%%%%%%%%%%%%%%%
\begin{abstract}\noindent
  Spin-rotators in the the few-MeV range require large trajectory deflection angles
  in order to change the spin direction appreciably. We describe a magnet system
  that accumulates adequately large bending angles in a small footprint and,
  at the same time, supports continuous-wave operation.
  % Such systems can even be incorporated in storage rings in order to increase
  % damping, albeit at higher energies.
\end{abstract}
%
% \begin{textblock}{5}(5,-6)\noindent\large
% \begin{flushright} JLAB-TN-26-0XX \end{flushright}
% \end{textblock}
%%%%%%%%%%%%%%%%%%%%%%%%%%%%%%%%%%%%%%%%%%%%%%%%%%%%%%%%%%%%%%%%%%%%%%%
%
\section{Introduction}
Occasionally, it is desirable to provide large values of integrated dipole fields
in an accelerator. Damping wigglers~\cite{DAMPWIG} used in some electron rings are
one example; they increase the power of the emitted radiation and thereby reduce
the damping time to reach the equilibrium emittance. Another example are spin rotators
that turn the polarization of a beam, often from the longitudinal direction to
the vertical, because that minimizes spin precession in subsequent dipole magnets.
At low energies, Wien filters~\cite{WIEN} can be used and at high energies interleaved
solenoids and dipoles~\cite{SPINROT} serve this purpose. In the mid-energy range
around a few MeV, however, the spin precession introduced by dipoles requires very
long or very strong dipoles. The reason is implicit in the Thomas-BMT equation~\cite{TBMT}.
It implies that the spin-precession angle is given by $\psi=\gamma a\phi$ where
$\phi$ is the deflection angle of the beam, $a=1.159\times 10^{-3}$ is the gyro-magnetic
anomaly of the electron and $\gamma$ is the energy of the beam in units of the
electron rest mass. At moderate energies the small values of both $a$ and $\gamma$
required very large values of $\phi$ to achieve even moderate values of $\psi$.
\par
Such a spin rotator is needed, for example, in order to rotate the polarization
of a positron beam~\cite{POSI}, because the capture of positrons after the production
target is likely based on accelerating cavities in order to exploit the benefits of
adiabatic damping. This helps to increase the number positrons within a small
transverse phase-space area. If the beam is pulsed, it is possible to accumulate
the required bending angle in a ring; just let the beam circulate the required
number of turns and extract it. Unfortunately, injection and extraction with
continuous beams is not feasible. One way out of this dilemma are multi-pi
rosetta magnets, which we discuss in the following.
\par
The Rosetta magnet is loosely inspired by the Rhodotron~\cite{RHOD}, based on a
coaxial accelerating cavity and turn-around dipole magnets that bend the beams
onto the next path though the cavity. This requires individual excitation of
subsequently traversed dipoles in order to provide the same bending angle for
beams with increasing energies. One important advantage of rhodotrons is that
they can accelerate continuous beams. We propose to operate such a magnet structure
at constant energy and, if still larger deflection angles are needed, assemble
multiple systems in a star-shaped formation.
\section{Rosetta magnet}
\label{sec:mag}
%
%..............................................
\begin{figure}[tb]
\begin{center}
\includegraphics[width=0.9\textwidth]{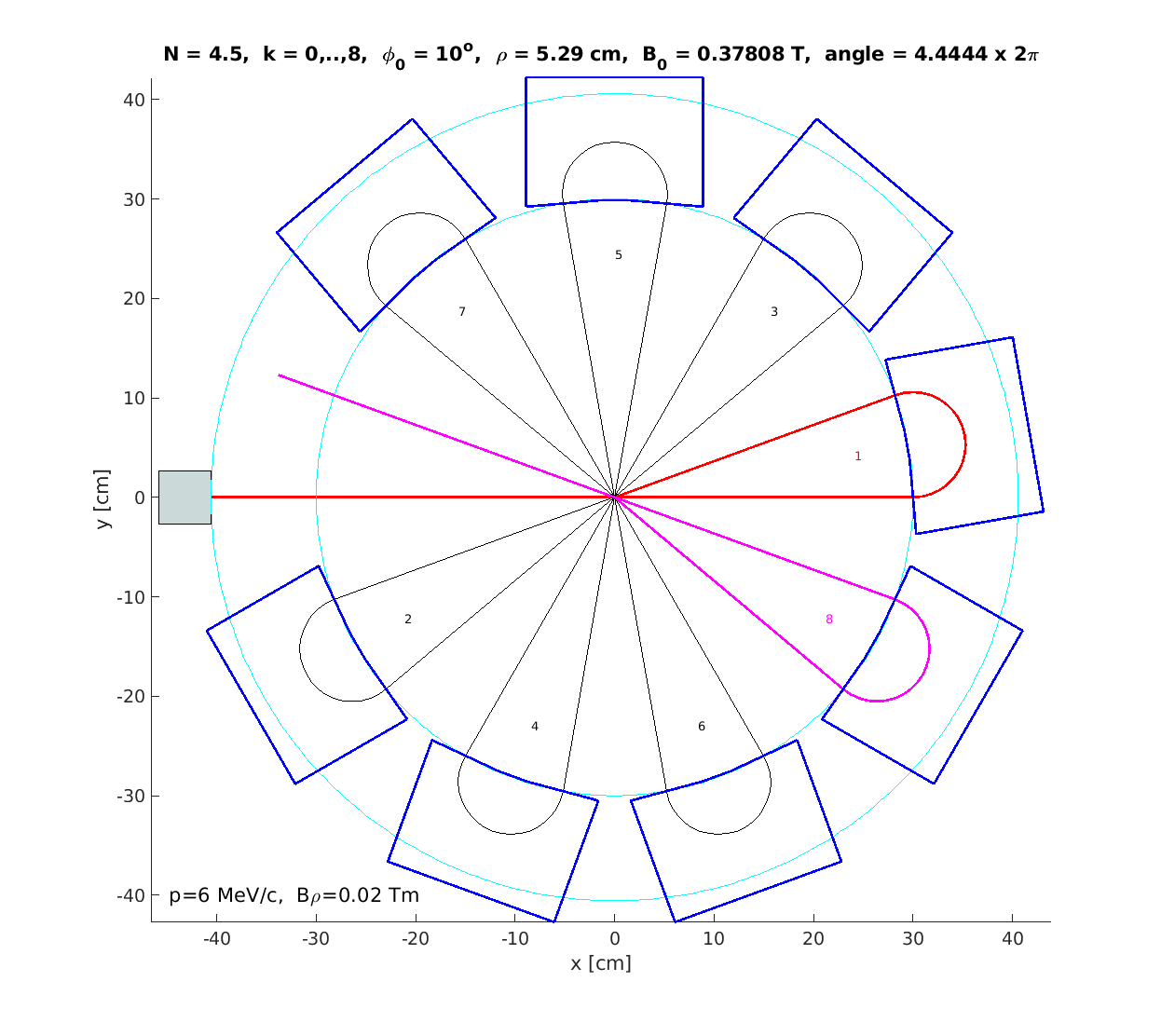}
\end{center}
\caption{\label{fig:ros}Rosetta magnet with eight ``leafs''. The parameters specified
  above the schematic refer to an electron beam with $p=6\,$MeV/c. They are explained
  in the text.}
\end{figure}
% ..............................................
Our rosetta magnet inherits the ability to handle continuous beams from the rhodotron.
It just omits the accelerating cavity and has all magnets excited equally. The geometry
for one rosetta is shown in Figure~\ref{fig:ros}. We see the beam following the
red line entering from the left, crossing the center of the magnet and moving towards
the blue magnet at the right. The small number indicates the sequence in which the
``leafs'' of the rosetta are traversed. The magnet turns the trajectory around such
that it crosses the center once again and is on its way towards leaf~2 on the left-hand
side. After being turned around it is on its way to leaf~3 and so forth. After traversing
seven leafs, the beam passes through the eighth leaf, shown in magenta, and exits the system
on the left, a little above the point where it entered.
\par
%..............................................
\begin{figure}[tb]
\begin{center}
\includegraphics[width=0.47\textwidth]{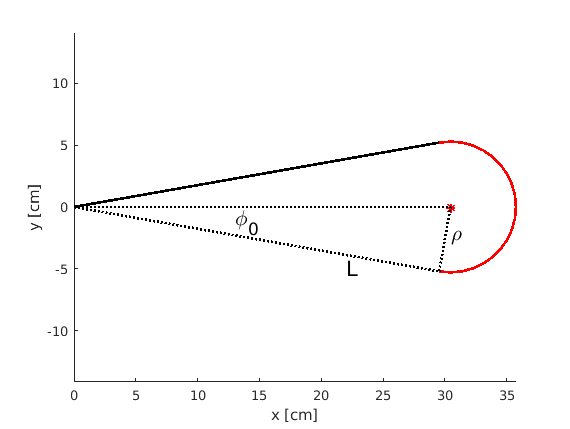}
\includegraphics[width=0.47\textwidth]{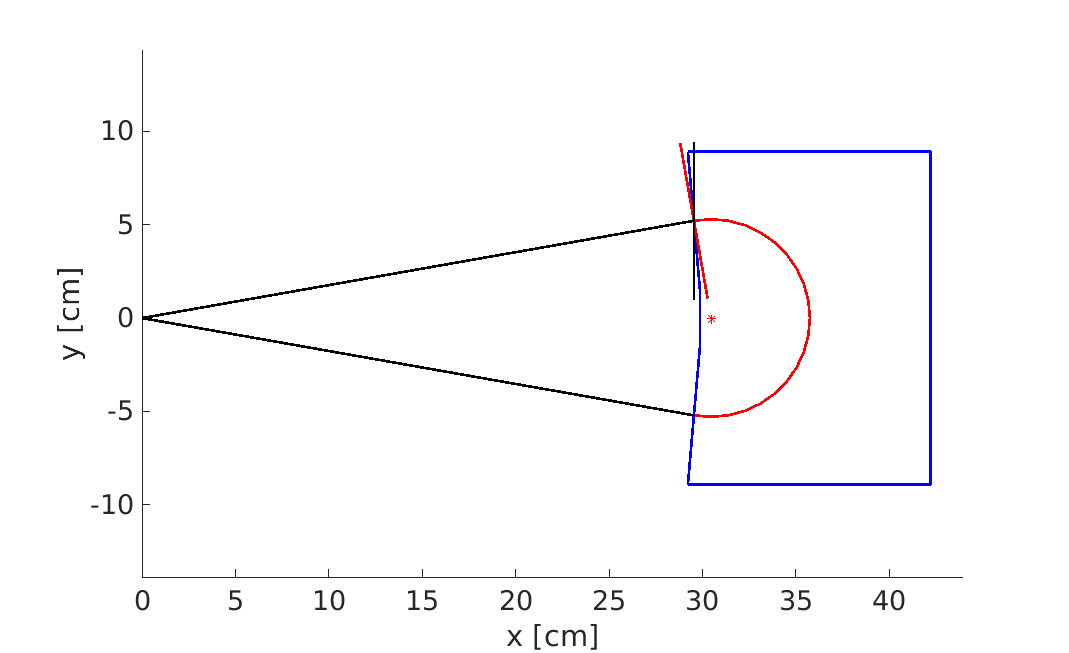}
\end{center}
\caption{\label{fig:rho}Left: the relation between length $L$, bending radius $\rho$,
  and half-angle $\phi_0$ of one leaf. Right: The trajectory of the beam with magnet
  superimposed. The straight lines at the upper part illustrate that the edge angle
  is half-way between the normal to the trajectory and the face of a rectangular dipole.}
\end{figure}
% ..............................................
The design of the system is governed by two parameters: the length $L$ from the center
to the entrance point into the magnet. In Figure~\ref{fig:ros} we chose $L=30\,$cm.
The other design parameter is the half-angle $\phi_0$ of one leaf, which we
equivalently can specify by the number $N$, which is related to $\phi_0$ via
$\phi_0=45^o/N$. We chose $N=4.5$. Both $\phi_0$ and $N$ are shown in the title
of the figure. The
number of magnets used for a particular value of $N$ is given by $k_{max}=2N-1$,
which leads to the eight magnets visible in Figure~\ref{fig:ros}. We also label
by a parameter $k=0,\dots,k_{max}-1$ the leafs in the sequence they are traversed.
Once $L$ and $\phi_0$ are specified, we can calculate the bending radius $\rho$
from $\rho=L\tan\phi_0$, which is illustrated on the left-hand plot in Figure~\ref{fig:rho}.
Moreover, simple geometric considerations imply that the magnet has to bend the
trajectory by $\theta=180^o+2\phi_0$, such that the length $s$ of the arc inside
the magnet is given by $s=\rho\theta$. Finally, for an electron momentum of $p=6\,$MeV/c
the rigidity is $(B\rho)=0.02\,$Tm and this requires a field of $B_0=0.378\,$T,
which is also indicated in the title of Figure~\ref{fig:ros}. The figure of merit
is the total accumulated angle $k_{max}(180^o+2\phi_0)=(2N-1)(180^o+45^o/N)$ which is
also shown in the title bar in units of $2\pi$. The value in Figure~\ref{fig:ros}
thus indicates that the total angle is 4.44 times a full turn and thus corresponds
to a bit over 4 turns in a storage ring.
\par
For an approximate electro-magnetic design of the magnet, we consider a C-type magnet with
a magnet gap of $h=2\,$cm. With the magnetic field $B_0$ we find that $NI=B_0h/2\mu_0
=3000\,$ampere-turns are needed to excite the upper and lower coils. Furthermore, we
will use a conductor with a cross section of $5\times 5\,$mm$^2$ with a water hole
having a diameter of 3\,mm. Such a conductor has a cross section of $A=18\,$mm$^2$. 
Assuming a current density of 7\,A/mm$^2$ (well below the 10\,A/mm$^2$ limit) each
conductor can carry up to 126\,A and we need 24 turns to reach the 3000\,ampere-turns.
Thus a sandwich of five layers with five turns each will suffice to power the magnet,
which looks feasible. In order to ensure that external connections are on the outside
of the coil, an even number of layers is preferable. For example, a $6\times 7$ coil
sandwich also reduces the current to about 72\,A, which will lead to more economical
power supplies without unduly increasing the overall size of the coil sandwich.
\par
The overall design can be easily scaled by selecting different parameters $L$ and $N$.
For example, larger values of $L$ cause $\rho$ to increase and that leads to lower
values of $B_0$. Similarly, lower values of $N$ lead to larger angles $\phi_0$ and that
also increases $\rho$ and lowers $B_0$.
\par
With the basic design parameters defined we now investigate the beam optical
properties of the system.
\section{Rosetta optics}
\label{sec:optics}
The high degree of symmetry implies that it is sufficient to analyze the
optical properties---the Twiss parameters---of a single leaf, which then
repeats $2N-1$ times. At the end we just have to match an incoming beam to
the Twiss parameters at the center of the system.
\par
%..............................................
\begin{figure}[tb]
\begin{center}
\includegraphics[width=0.7\textwidth]{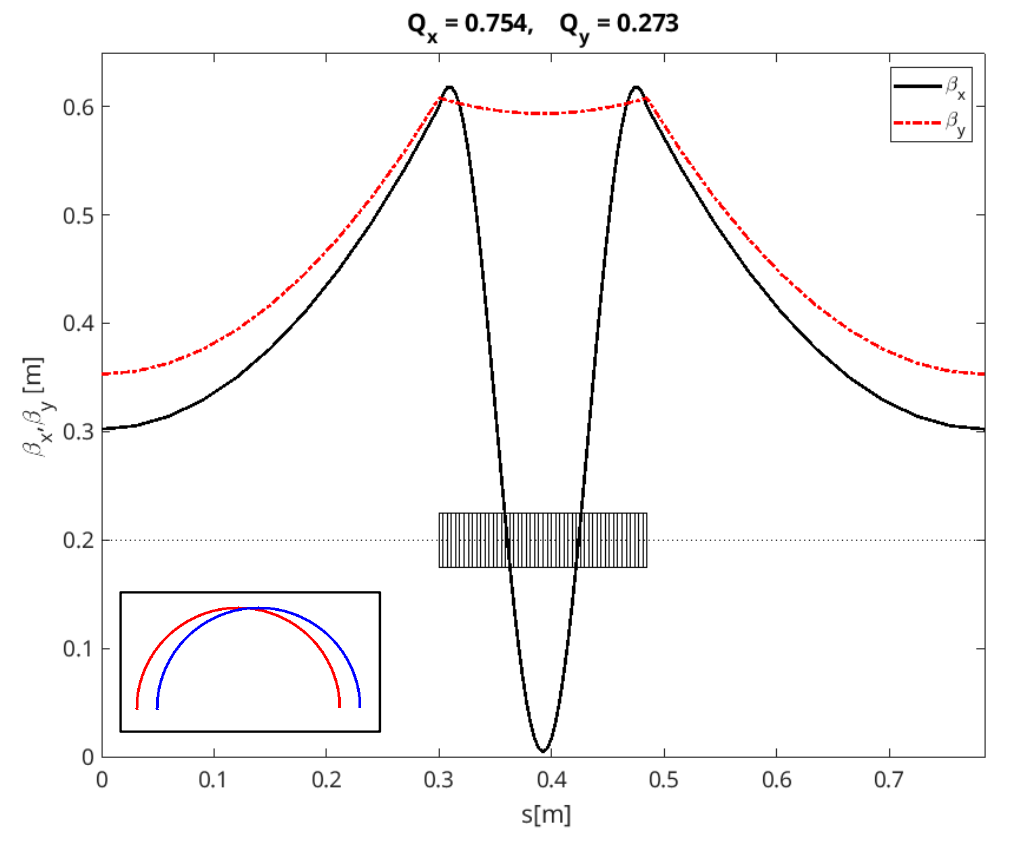}
\end{center}
\caption{\label{fig:opt}Horizontal (black) and vertical (red) beta functions
  for one leaf with the phase advances in units of $2\pi$ (the tunes) are
  shown above the plot. The insert illustrates the reason for the focus in
  the  center of the magnet.}
\end{figure}
% ..............................................
The turn-around magnet is the dominant component that defines the beam
optical properties of a leaf where the main factor is the so-called
weak focusing in the horizontal plane. If we make the dipole a sector
dipole with normally incident trajectories, there would be no vertical
focusing. We therefore introduce pole-face rotation angles with a
magnitude of $\phi_0/2$ which is half-way between normal incident and
a configuration that resembles a rectangular dipole. A similar configuration
is already used in rhodotrons~\cite{RHOD2}, which additionally add
vertical focusing by shaping the pole face. In this report and for
simplicity, we only use $\phi_0/2$ which provides approximately equal
focusing in the horizontal and the vertical plane. We thus model the
magnet as a sector dipole of length $s=\rho\theta$, already defined
above with pole-face angles $\phi_0/2$. On either side of the magnet
a drift space of length $L$ completes the description of one leaf.
We use the software described in~\cite{VZAP} to model the beam optics
of this system.
\par
Figure~\ref{fig:opt} shows the corresponding beta functions. We see
that both horizontal and vertical beta functions outside the dipole
are close, indicating  that the focusing is equally divided among the
two planes. The very sharp minimum of the horizontal beta function
has a geometric origin that is illustrated in the small insert at the
bottom left in Figure~\ref{fig:opt}. Two horizontally displaced
trajectories, shown in read and blue, have the same radius of
curvature in a homogeneous magnetic field. At the end points they
are displaced, but half-way in between they cross, which corresponds
to a focal point. This is the origin of the focal point that the
horizontal beta function shows. It is also the origin of a $180^o$
phase shift that accounts for the difference in the ``tunes''
shown above the plot.
\par
The magnitude of the beta functions at the start and end of the plot
is approximately equal to the design parameter $L$. Thus it can be
adjusted to help matching into the beam lines that attach to the entrance
and exit of the system. We point out that the tunes, or equivalently,
the phase advance for one leaf is only weakly affected by changing $L$.
\section{Combining and matching}
%
%..............................................
\begin{figure}[tb]
\begin{center}
\includegraphics[width=0.9\textwidth]{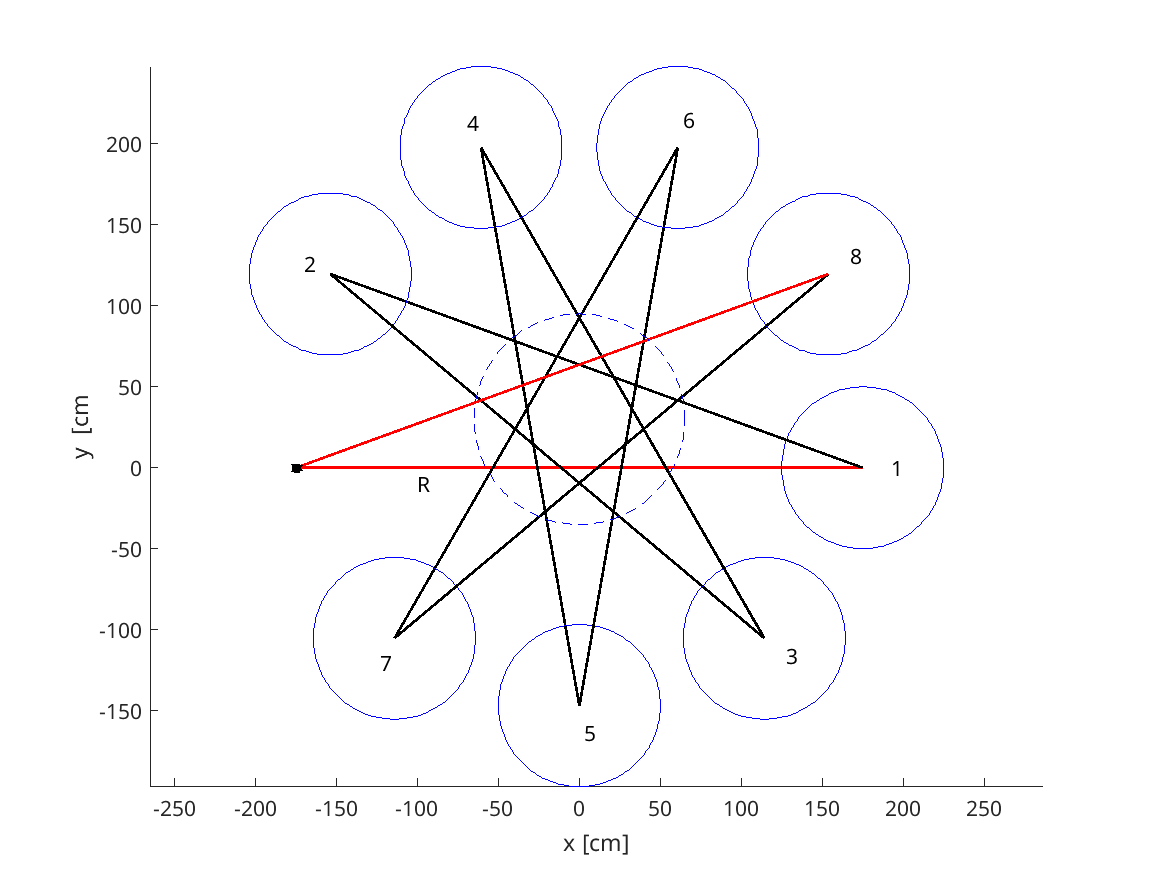}
\end{center}
\caption{\label{fig:star}Combination of eight rosettas (blue circles) to
  a star-shaped system. The beam enters the system along the red horizontal
  line toward rosetta~1, leaves it with a 20\,degree angle and moves to
  rosetta~2. The pattern repeats until the beam exits rosetta~8 along the
  diagonal red line. the dashed circle in the center denotes the exclusion
  zone to place quadrupole magnets.}
\end{figure}
%..............................................
The integrated bending angle of the rosetta magnet from Figure~\ref{fig:ros}
corresponds to 4.44 turns, which only turns the polarization by
$\Delta\psi = 4.44\times 360^o \times \gamma a = 21.8^o$. In order to
rotate the polarization by $90^o$ we therefore need at least 4 rosetta
magnets. An this requires that we match one rosetta magnet to another
to create a sequence of them.
\par
First we note that the direction of the beam when exiting a rosetta is
deflected by $180^o-2\phi_0=160^0$, which causes a sharp needle-like
point in the center of the rosetta. If we now follow the direction of the exiting beam and, after
some distance, pass it through a second rosetta, and then continue in
the same fashion, we see from Figure~\ref{fig:star} that the trajectory
follows a star with nine rays. The blue circles indicate the size of
the rosetta in Figure~\ref{fig:ros} to illustrate the size of the entire
system. It also helps us to prevent collisions between adjacent rosettas.
In Figure~\ref{fig:star} we use $R=350\,$cm. The numbers in the circles
indicate the sequence in which the rosettas are traversed when the
beam enters along the horizontal direction from the left at the point
marked by the black asterisk. After traversing rosetta number 8
it will exit towards the lower left direction. In the complete setup,
the beam will traverse eight rosettas, but we can easily omit
rosettas five through eight and the beam will exit the structure
at the point of rosetta number 5 towards bottom of the graph. In this
case we have constructed a structure that will add the angles of the
four rosettas, because all deflection angles have the same sense. This
will therefore provide a spin rotation angle of $4\times 4.44\times 360^o
\gamma a = 87.3^o$. Importantly, this system will work without injection
or extraction and will therefore be able to operate with continuous
beams. 
\par
%..............................................
\begin{figure}[tb]
\begin{center}
  \includegraphics[width=0.44\textwidth]{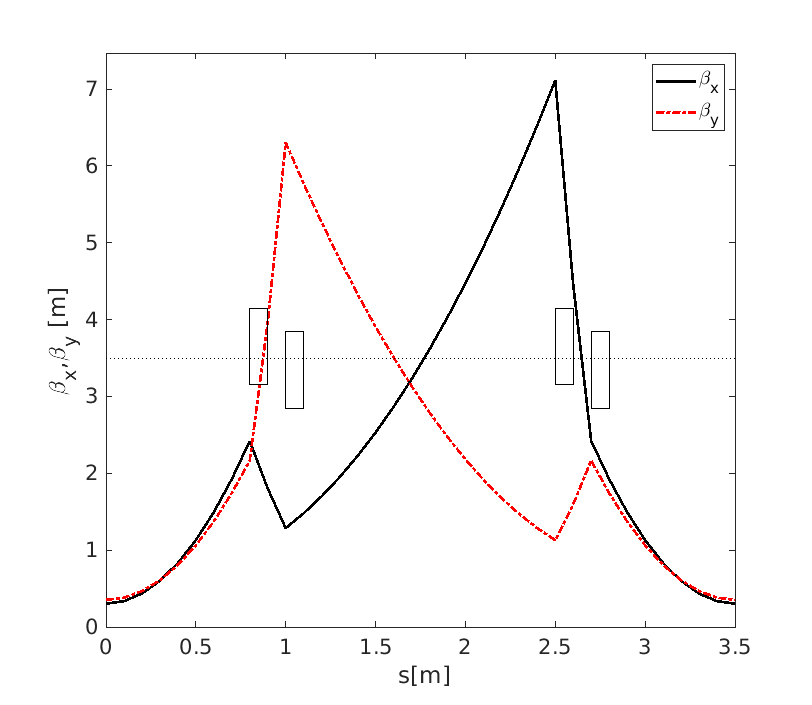}
  \includegraphics[width=0.54\textwidth]{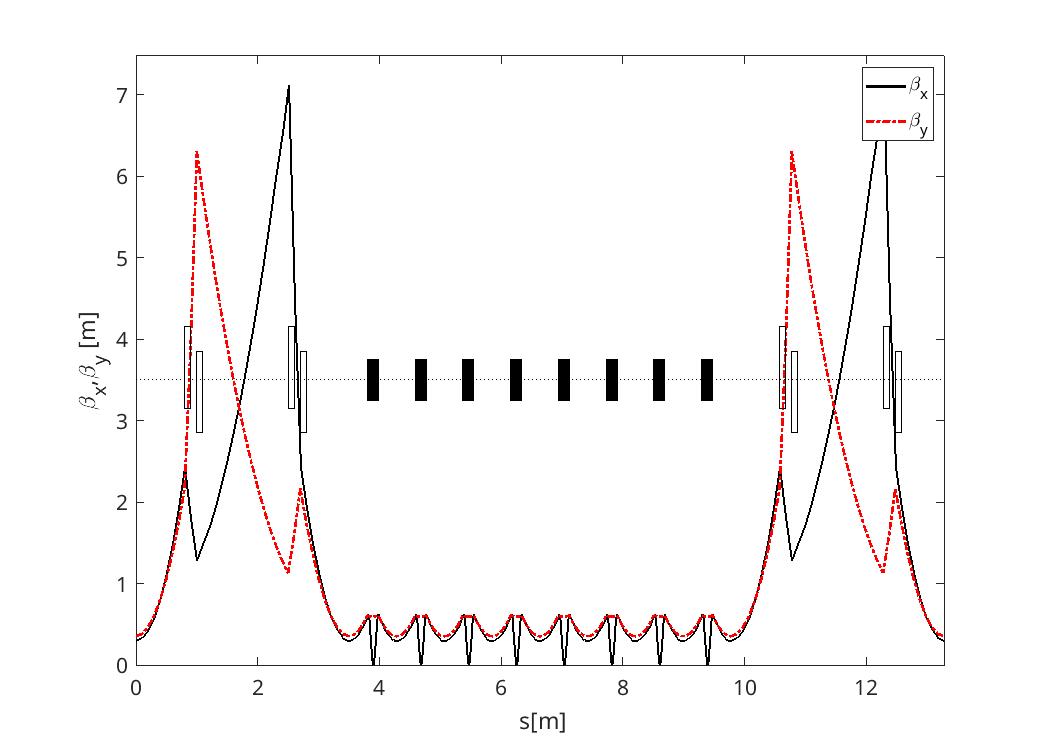}
\end{center}
\caption{\label{fig:straight}Left: Horizontal (black) and vertical (red)
  beta functions for the straight section between two rosettas. The magnets
  are superimposed. Right: Beta functions for a rosetta inbetween two
  straight sections.}
\end{figure}
%..............................................
It remains, however, to match the beta functions at the center of two
rosettas to each other.  But we have 350\,cm space to do so. We only
have to avoid placing magnets in the region, indicated by the dashed
circle, where the beams cross. We know that the natural beta functions
at the center of each rosetta are the starting values shown in
Figure~\ref{fig:opt}. They are $\beta_x=0.302\,$m and $\beta_y=0.353\,$m
with $\alpha_x=\alpha_y=0$. Since we want to avoid placing quadrupoles
inside the central region, we chose a doublet configuration, which
leaves sufficient space between the magnets while still being able
to focus down to the rather small beta functions at the entrance
to the rosetta. We also chose to use thin-lens quadrupoles to simplify
the analysis.
\par
Exploring different configurations of magnet polarities, we arrived at
the configuration shown on the left-hand side in Figure~\ref{fig:straight}.
The magnitude of the four focal lengths are around $f\approx\pm0.4\,$m. We further
observe that the horizontal (black) and vertical (red) beta functions show
a distinct crossing at the center, which will help us later to match the
star-shaped configuration to a generic FODO cell of a generic injection
beamline. The right-hand plot in Figure~\ref{fig:straight} shows the
connection of the doublet cells in the straight sections of the star to a
rosetta with the distinct eight leafs, whose beta function is shown on
a larger scale in Figure~\ref{fig:opt}. The straight sections nicely match
into the rosetta and out again, connecting to the next rosetta, which,
however, is not shown in the figure.
\par
%..............................................
\begin{figure}[tb]
\begin{center}
  \includegraphics[width=0.85\textwidth]{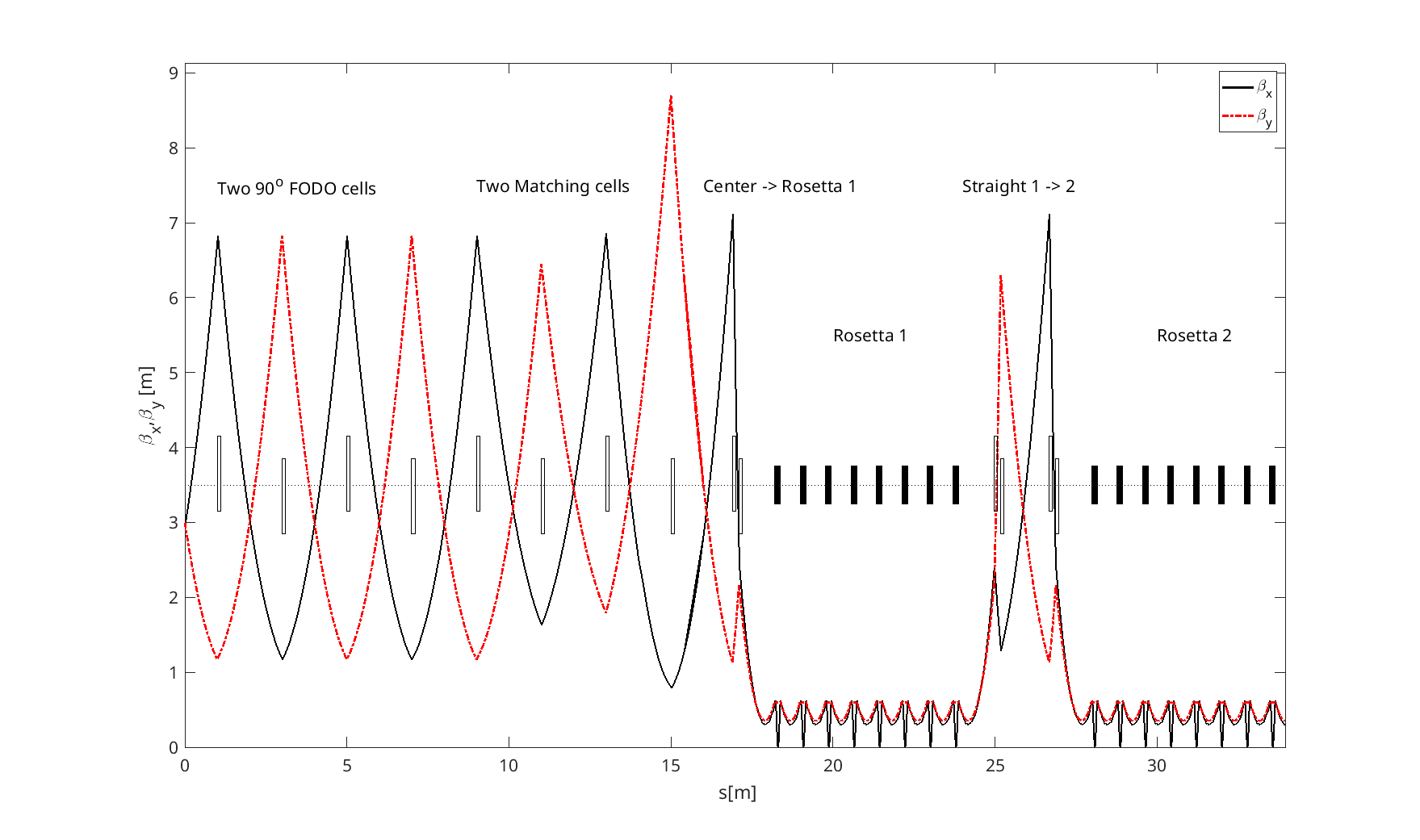}
\end{center}
\caption{\label{fig:match}Beta functions for the matching section to match a generic
  $90^o$ FODO beamline to the star-like system shown in Figure~\ref{fig:star}. The
  respective functional blocks are indicated in the plot.}
\end{figure}
%..............................................
Finally, we have to match to a generic beamline, we chose a thin-lens FODO
cell with a length of 4\,m and a phase advance per cell of $90^o$ in both
planes. We then match this cell to the center of the star with the help of
two cells with individually variable matching quadrupoles to adjust $\beta_x$,
$\beta_y$, $\alpha_x$, and $\alpha_y$ to the values in the center of the
straight where the beta functions cross on the left-hand plot in
Figure~\ref{fig:straight}. The excitation of the four matching quadrupoles
differs only a little from the regular settings for the $90^0$ lattice. 
Figure~\ref{fig:match} shows the entire injection from the FODO line to the
first rosetta, through the first straight and on to the second rosetta.
We see that the beta functions match very well. In particular, the dual
purpose of the doublet lattice in the straights becomes apparent. First,
it acts like a final focus system and demagnifies the incoming beta functions
by more than an order of magnitude to fit the small beta functions in the 
rosetta. Second, it connects one rosetta to the next with only four quadrupoles
and, at the same time, leaves sufficient space to prevent hampering the beam crossings
at the center of the star. Finally we point out that the extraction from the
star can be implemented as the mirror image of the injection.
%
% \section{Damping}
% %
% The ability to match the tight focusing rosetta to a system with much larger beta
% functions can be used to insert a rosetta-based system in a ring in order to
% increase the amount of synchrotron radiation and thereby reduce the damping
% time without unduly increasing the footprint of the accelerator. Here we consider
% a generic 1\,GeV electron damping ring with a circumference of 100\,m that uses
% a $90^o$ FODO lattice. The two halves of the ring we pull apart and insert a
% rosetta-based star
% 
\section{Conclusions and Outlook}
\label{sec:conc}
%
%..............................................
\begin{figure}[p]
\begin{center}
  \includegraphics[width=0.47\textwidth]{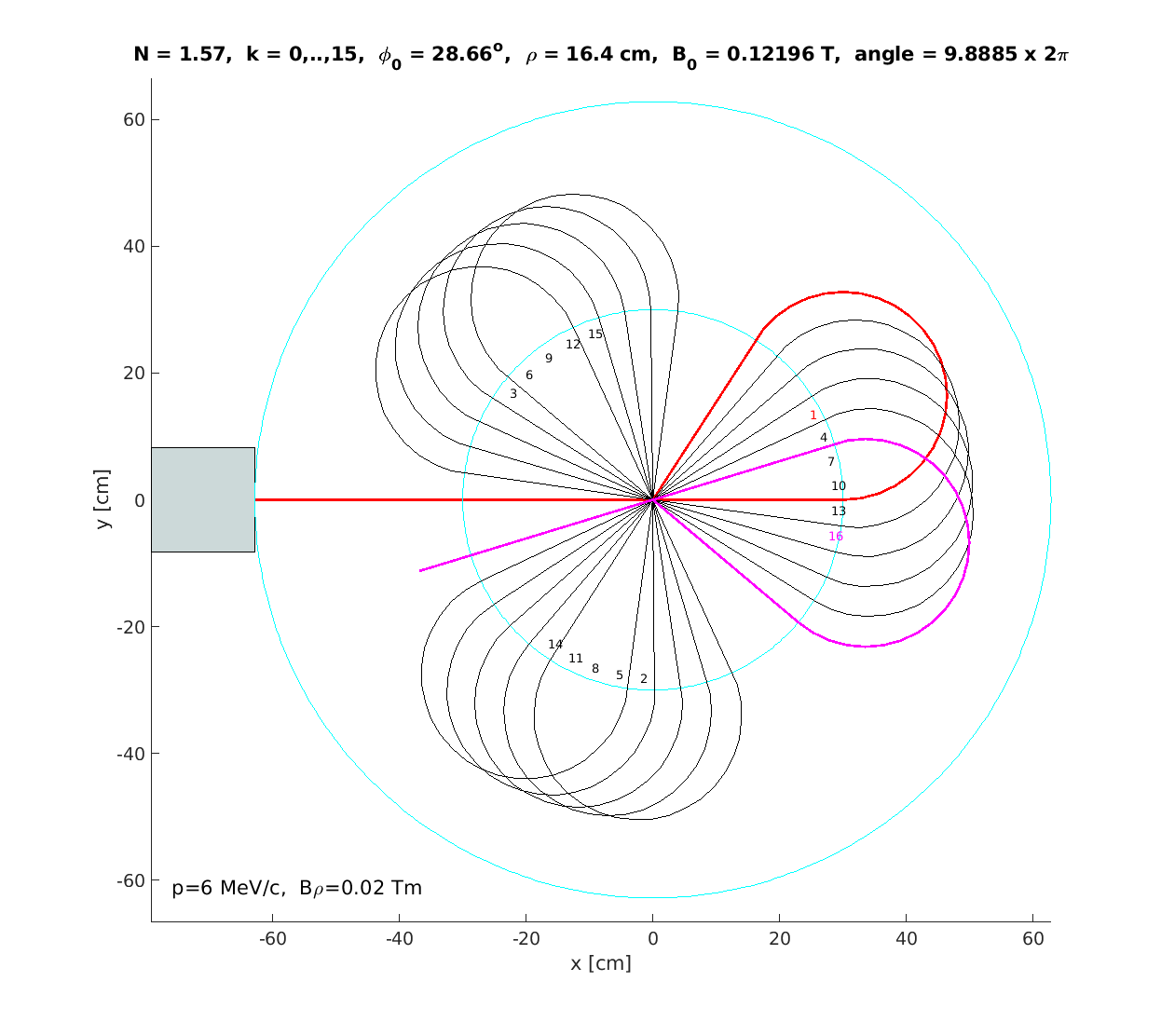}
  \includegraphics[width=0.47\textwidth]{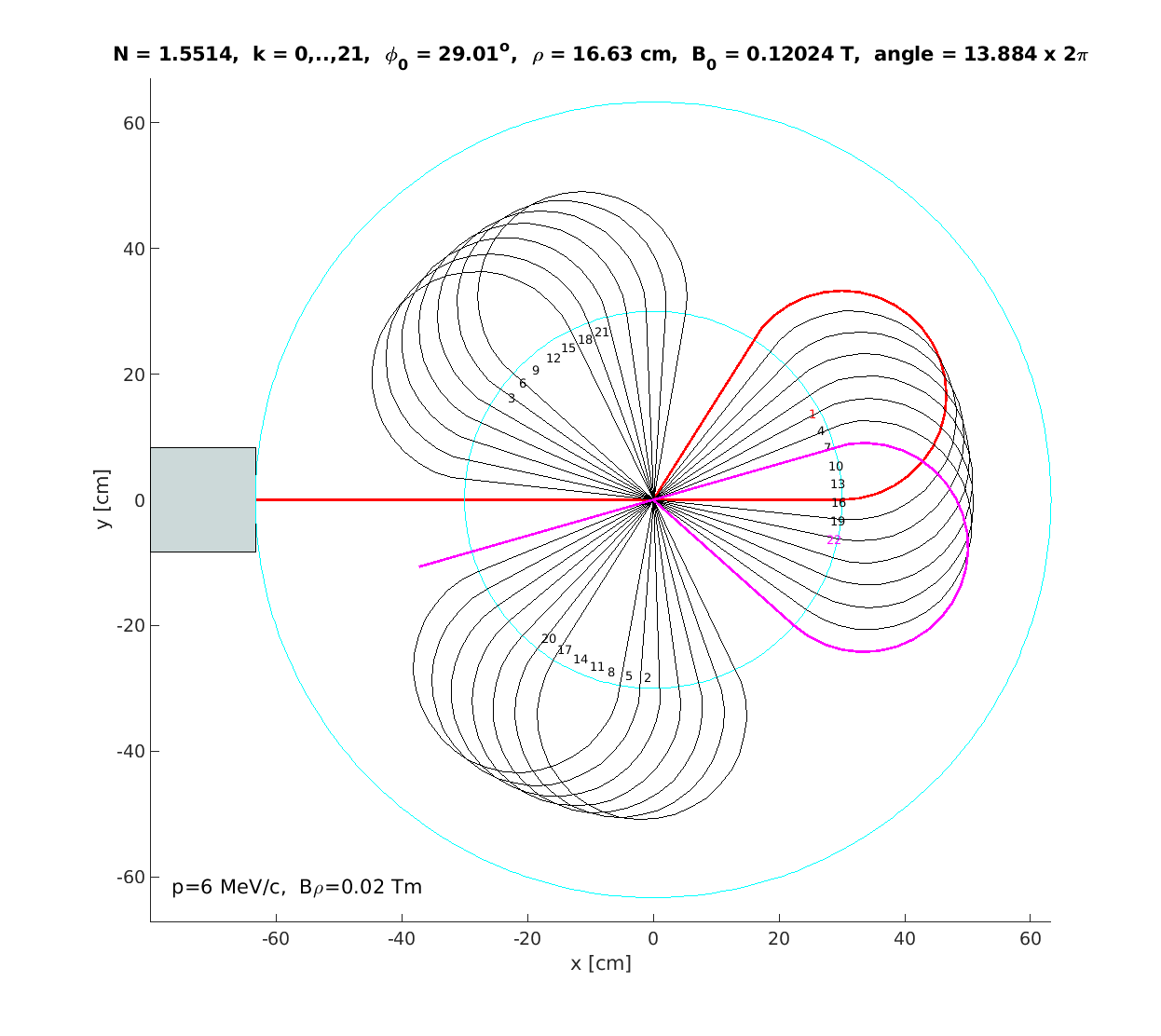}
  \includegraphics[width=0.47\textwidth]{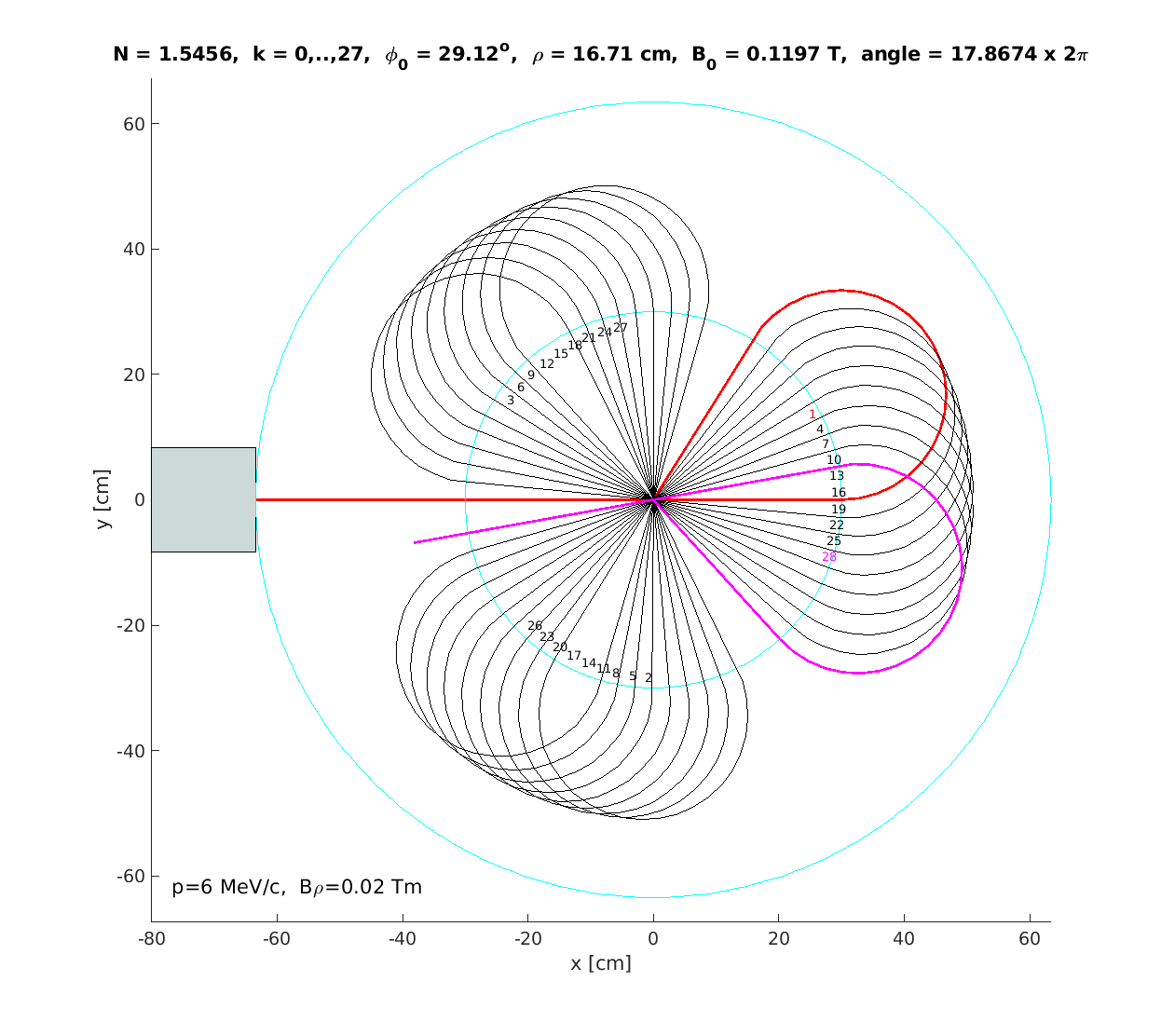}
  \includegraphics[width=0.47\textwidth]{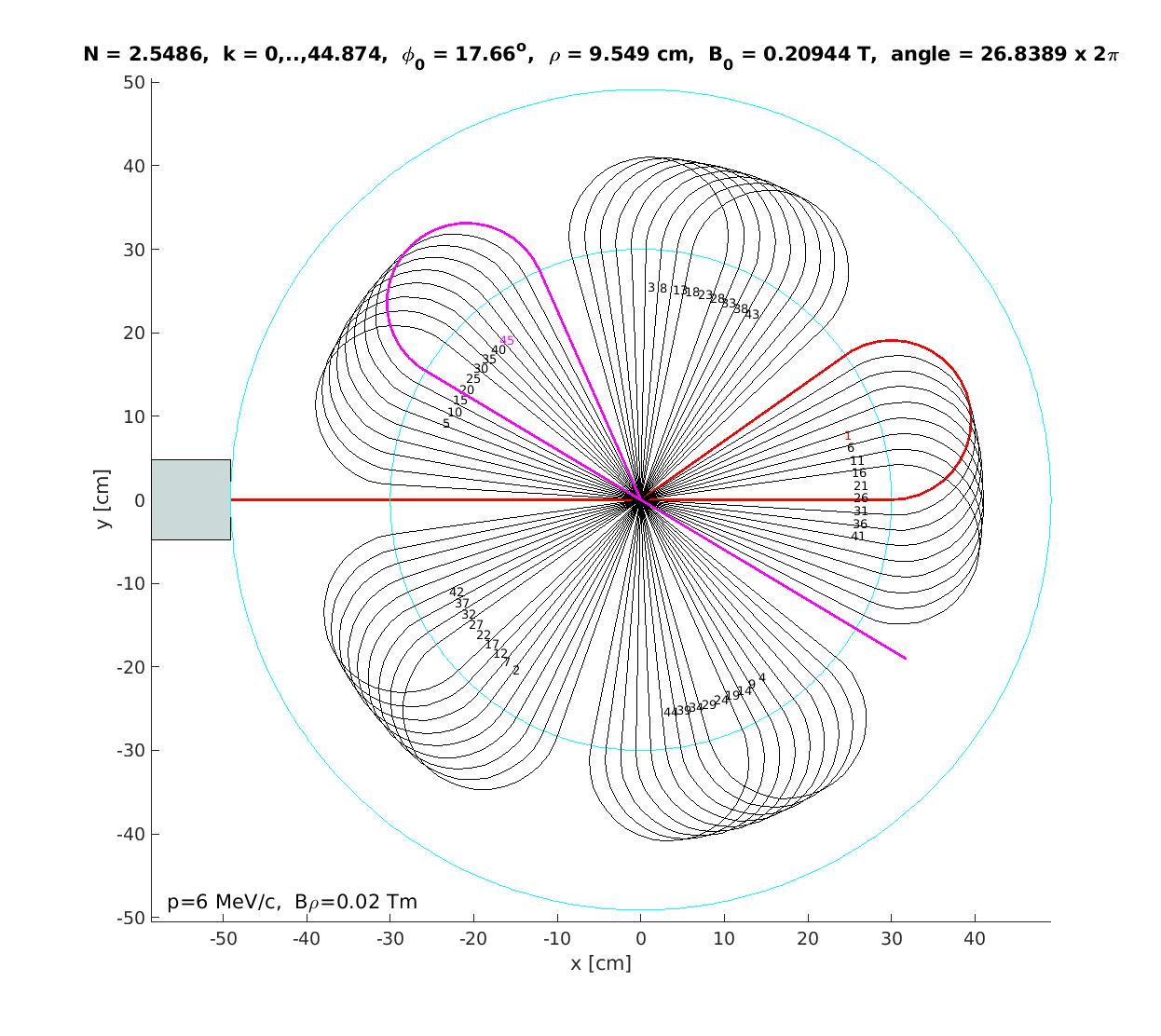}
\end{center}
\caption{\label{fig:mros}Four rosetta configuration where the beam traverses the
  same magnet multiple times. The turn count increases from 9.8 on the top left
  to 28.8 on the bootom right. All parameters are given in the title bar of each
  plot.}
\end{figure}
%..............................................
We described a system of magnets, capable of supporting continuous-wave operation
with beam, that accumulate large deflection angles in limited floor space. The
basic building block is one leaf of the rosetta structure, shown in Figure~\ref{fig:ros}
that is inspired by the geometry of a rhodotron. Each of the turn-around magnets
that constitute the rosetta deflects the beam by more than $\Delta\phi=180^o$ and the
nominally ``weak'' focusing is actually rather strong and causes the beta functions,
shown in Figure~\ref{fig:opt}, to be rather small. It is, however, possible to
combine multiple rosettas in a star-like structure using the doublet lattice shown
in in Figure~\ref{fig:straight}. Even matching to a generic FODO lattice in order
to inject and extract is straightforward, as is witnessed in Figure~\ref{fig:match}.
\par
Each of the rosettas accumulates a deflection angle of $\phi=4.44\times 2\pi$ and
will rotate the polarization by $\psi=\gamma a \phi = 21.8^o$ for a beam with
6\,MeV/c momentum. Thus, in order to rotate by $90^o$, we need to combine four
rosettas. The star-shaped system, shown in Figure~\ref{fig:star}, can combine
up to eight rosettas, though four would suffice to approximately achieve the desired
$90^o$ rotation. A small additional section of deflection magnets might be needed
to exactly obtain $90^o$ but we did not pursue this further. Optionally, we can
chose to operate the rosetta at a different momentum and associate $\gamma$ and
thereby exactly reach $90^o$ spin rotation.
\par
The rosetta shown in Figure~\ref{fig:ros} accumulates 4.44 turns in a compact space.
One might wonder whether it is possible to run through the same rosetta multiple times
with a similar magnet system. The problem in that case is focusing in the vertical
plane, because we rely on the vertical focusing from the magnet entrance and exit
and mixing too many orbits will make that very difficult.
\par
In special situations, however, multiple orbits organize themselves in such a way
that we can still use equal edge angles on one side of the magnet and likewise
the opposite edge angle on the other side. See for example, the top left plot in
Figure~\ref{fig:mros} which displays three groups of leaves. Each group clearly
separates the orbits entering the magnet (on the clock-wise side) from those
exiting the magnet (on the counter-clock-wise side). In this case, we would
therefore need three magnets with inwardly inclined pole faces, similar to those
shown in Figure~\ref{fig:ros}. The configuration is characterized by $L=0.3\,$m and
$N=8/5-0.03=1.5700$ with other parameters displayed in the title of the plot. The
benefits of this configuration are substantial, because the accumulated deflection
angle becomes 9.88 turns while the field in the dipole with 0.122\,T is much weaker
than that of the original rosetta from Figure~\ref{fig:ros}.
\par
Slightly changing $N$ to $N=11/7-0.02=1.5514$ produces the configuration in the top right
plot. It accumulates 13.8 turns. The relative position of the orbits remains
equal to that shown on the top left plot. The value of $N$ in the bottom left
plot is $N=14/9-0.01$ which increases the accumulated angle to 17.87 turns.
The bottom right shows a configuration with $N=23/9-0.07$ leading to an
accumulated angle of 28.84 turns.
\par
All of theses configuration have well-separated in and out sides of the respective
magnets, which allows us to utilize edge focusing to focus in the vertical plane.
We note that the separation between the sides with incoming and outgoing trajectories
becomes progressively narrower the larger the turn-count becomes which will become
more difficult to manufacture and control tolerances. We do not explore these
advanced options further, but defer their analysis to a later report.
\section*{Acknowledgements}
Fruitful discussion with my colleague Jay Benesh are gratefully acknowledged.
Notice: This work was produced in part by SURATech, LLC under Contract No. 89243126CSC000213
with the U.S. Department of Energy. Publisher acknowledges the U.S. Government license and
provides public access under the DOE Public Access Plan (\url{http://energy.gov/downloads/doe-public-access-plan}).
%
%%%%%%%%%%%%%%%%%%%%%%%%%%%%%%%%%%%%%%%%%%%%%%%%%%%%%%%%%%%%%%%%%%%%%%%
%
\bibliographystyle{plain}

\end{document}